\documentstyle[amssym,emulateapj,epsf]{article}

\slugcomment{}

\lefthead{Treu \& Stiavelli}
\righthead{High redshift ellipticals}

\begin{document}

\title{A NICMOS search for high redshift elliptical galaxy candidates \altaffilmark{1}}

\author{Tommaso Treu\altaffilmark{2} and Massimo Stiavelli
\altaffilmark{3,4}} \affil{Space Telescope Science Institute, 3700 San
Martin Dr., 21218 MD, U.S.A.; treu@stsci.edu, mstiavel@stsci.edu}

\altaffiltext{1}{Based on observations collected at ESO, La Silla, at
the CTIO, NOAO, which is operated by AURA, under cooperative agreement
with the NSF, and with the NASA/ESA HST, obtained at the STScI, which
is operated by AURA, under NASA contract NAS5-26555.}
\altaffiltext{2}{Scuola Normale Superiore, Piazza dei Cavalieri 7,
I56126, Pisa, Italy}
\altaffiltext{3}{On leave from Scuola Normale Superiore, Piazza dei
Cavalieri 7, I56126, Pisa, Italy}
\altaffiltext{4}{On assignment from the Space Science Department of
the European Space Agency}

\begin{abstract}

We have collected optical follow-up observations from the ground and
the Hubble Space Telescope (HST) for 13.74 $\sq '$ of archival images
taken with HST Near Infrared Camera and Multi Object Spectrograph
(NICMOS).  We use two criteria to select E/S0 galaxy
candidates at $z \ga 1$ and $z \ga 1.5$ (HizECs) based on colors and
infrared morphology. The observed surface density of HizECs is
significantly smaller than what is predicted assuming the local
luminosity function (LF) of E/S0, constant comoving density, and pure
passive evolution with a high redshift of formation of the stellar
populations ($z \ga 5$). On the other hand, assuming a low redshift
($z \la 2$) of formation, not enough HizECs are predicted. The data
are very well matched assuming that a substantial fraction of
present-day E/S0s (10--66\%, depending on cosmology and on the
local LF of E/S0s) formed at $ z = 3$ or higher
and then evolved passively. The rest of the current E/S0 population
may have formed at lower redshift, or may have been rejected by the
selection criteria because of interactions or more recent episodes of
star-formation (disturbed morphology and/or bluer colors).

\end{abstract}

\keywords{galaxies: elliptical and lenticular, cD --- evolution --- formation ---; cosmology: observations; infrared: galaxies}

\section{Introduction}

The formation of elliptical galaxies is still not understood. In a
schematic view, all the different models can be summarized within two
simple paradigms: {\it i)} the monolithic collapse, in which
ellipticals form at high redshift in short bursts of star
formation; and {\it ii)} the merging scenario in which they form at
lower redshift, by merger of disk galaxies.

In recent years, many observational programs have started to provide
useful constraints on theoretical models.  Studies such as MORPHS, a
Hubble Space Telescope (HST) survey of intermediate redshift clusters
(up to $z \approx 0.5$; e.g. \cite{MORPH}), or the investigation of
the Fundamental Plane of ellipticals at intermediate redshift in
clusters (\cite{DFKI98}) and in the field (\cite{T99a}), indicate that
the bulk of star formation occurred at $z\ga1$.  However, low redshift
($z \la 1$) diagnostics are mostly sensitive to the epoch when stars
were formed rather than to that when galaxies were assembled (see
however Lilly et al. 1995; Kauffmann, Charlot \& White
1996). Unfortunately, it is in their predictions on the latter that
galaxy formation models differ the most.  Thus, the most direct and
uncontroversial way to break the degeneracy is to push the
investigation to higher and higher redshift, trying to put limits on
the number of already assembled (proto-)ellipticals.
The Near Infrared Camera and Multi Object Spectrograph (NICMOS)
parallel images can provide a decisive breakthrough onto this issue.
In fact, high redshift ellipticals appear extremely red in their
optical-infrared colors (e.g. \cite{SP}), and morphological
information is the best way, short of spectroscopic identification, to
distinguish between high redshift old stellar populations and
reddened dusty starbursts (e.g., Graham \& Dey 1996; \cite{S99},
hereafter S99).

By using the NICMOS parallel images combined with deep optical images,
both from ground based telescopes and with the Wide Field and
Planetary Camera 2 (WFPC2) on HST, we have undertaken a survey
(13.74~$\sq '$) to determine the number of ellipticals already
assembled at $z\ga1$. The aims of the survey, the selection criteria
and the sample are described in Sec.~\ref{sec:survey}. The building of
the catalog is briefly discussed in Sec.~\ref{sec:data}. The results
and conclusions are discussed in Sec.~\ref{sec:res}. A more detailed
description of the data reduction, the complete catalog of the
selected sources, together with images, surface photometry and
photometric redshifts, will be given elsewhere (Paper II, in
preparation. See also http://icarus.stsci.edu/$\sim$treu/HizECs).

AB magnitudes (\cite{Oke}) are used throughout the paper. The HST
filters are indicated as: H$_{16}$=F160W and J$_{11}$=F110W
(NICMOS); I$_{814}$=F814W and V$_{606}$=F606W (WFPC2).  H$_0$ is
assumed to be 100 $h$ km/s/Mpc. The present value of the matter
density and the cosmological constant are indicated by
$\Omega$ and $\Omega_{\Lambda}$, respectively.

\section{The survey}

\label{sec:survey}

Due to the effect of redshift and the lack of UV emission in old
stellar populations, ellipticals at $z$=1--2 are relatively bright in
the IR, but they are faint in the optical. For example a single burst
model of solar metallicity (\cite{BC93}; GISSEL 96 version; hereafter
BC96) with Salpeter IMF and cutoffs at 0.1 and 100 M$_{\odot}$,
normalized to a luminosity of M$_{V}$=-21.0 at 15 Gyrs (roughly
M$_{\star}$), at 2 Gyrs and $z$=1.5, has H$_{160}$=21, J$_{110}$=22,
I$_{814}$=23.7, R = 25.3, V$_{606}$ = 26, assuming $\Omega=0.3$ and
$\Omega_{\Lambda}=0.7$.  For this reason, high redshift ellipticals
can constitute a significant fraction of the so-called Extremely Red
Objects (EROs).  Unfortunately, even the best color-based
classification suffers from the degeneracy between old stellar
populations and dusty starbursts. High--resolution imaging holds the
key to classify these objects properly since many EROs are so faint
that their continuum cannot be detected spectroscopically, even with
8-10 meter telescopes.  For example, the nature of the two best known
EROs with spectroscopic information is fully revealed by their
morphology: {\it i)} HR10, a dusty starburst at $z$=1.44
(\cite{DG99,Cima}) is distorted in the HST images, while {\it ii)}
LBS53W091 is smooth and shows a perfectly regular $r^{1/4}$ light
profile (Dunlop 1998). From the statistics of EROs we know that the
density of high redshift ellipticals ($z\ga1$) must be of the order of
1$/\sq'$, the precise value depending on the definition and the
magnitude limit (e.g., \cite{CADIS}; \cite{Calred}; \cite{Yan}).

For all these reasons, the essential ingredients for a successful
search for high redshift ellipticals are: 1) High resolution IR
imaging. The morphology is best studied in the IR, where sources are
brighter and the surface brightness is a better tracer of the mass
distribution and is less affected than the optical by, e.g., a dwarf
galaxy companion or an otherwise negligible amount of star
formation. 2) Medium deep IR imaging, as the galaxies we are looking
for are expected to be in the range H$_{16}$=20-24. Deeper images
would probe a wider range of the luminosity function, but it would be
difficult to measure the flux of the optical counterparts. 3) The
largest possible area.

\subsection{Selection Criteria}

In order to count high redshift elliptical candidates
(HizECs) properly we need a definition excluding as much as possible
galactic stars and dusty star-forming objects. We require the object
to be: {\it i)} clearly resolved, i.e. FWHM $\ga0\farcs3$, hence
having a scale length $\ga$ 3 kpc, depending on the redshift and the
cosmological parameters; {\it ii)} to appear roundish and regular,
with no spikes, tails or knots; {\it iii)} to have colors consistent
with a BC96 model of old stellar populations ($\ga 1$ Gyr) at $z
\ga1$.  In order to quantify the third criterion, we have computed the
observed colors of a range of models as a function of redshift. For a
given cosmology and metallicity we computed the colors of a passively
evolving stellar population formed at redshift $z_f$=10, and of a 1
Gyr old single burst stellar population.  These two cases represent
the red and blue extrema of the objects we want to select and
therefore bracket the possible cases.  
Based on the computed colors we define a broad and a narrow
selection criteria. The broader criterion, tuned to collect
all the elliptical candidates at $z\ga1$, is: I$_{814}$-H$_{16}>1.8$,
I-H$_{16}>$1.9, R-H$_{16}>$3.2, and V$_{606}$-H$_{16}>$3.8. The more
stringent criterion, tuned to $z\ga1.5$ is defined by
I$_{814}$-H$_{16}>2.7$, I-H$_{16}>2.8$, R-H$_{16}>4.0$,
V$_{606}$-H$_{16}>4.5$. As the J$_{11}$-H$_{16}$ color is much
smaller, and therefore more affected by photometric (of order of
0.2--0.3 mags) and modeling errors, we do not include it into the
color selection criterion, but we shall use it, when available, as a
consistency check.

Extremely high redshift galaxies can appear very red when Ly$\alpha$
Forest absorption (\cite{MP1}) starts to affect the V$_{606}$ and R
filters. Nevertheless, the colors of extremely high redshift galaxies
can easily be distinguished from those of an old stellar population at
$z$=1--2 because the cosmic transmission falls bluewards of 1215 \AA\
much more sharply than the expected HizECs Spectral Energy
Distribution (\cite{MP}). Furthermore, galaxies at $z \ga6$ would need
to be enormously bright in order to be detectable at H$\sim24$ (Chen,
Lanzetta \& Pascarelle, 1999; \cite{z5p6}).  Therefore we assume the
contamination to our sample of HizECs by extremely high redshift
galaxies to be negligible.

\subsection{The sample}

In order to survey the largest possible area we collected data from
different sources: 1) The pointings of the HST NICMOS parallel data
taken with NIC3 through F160W were cross-correlated with the entire
WFPC2 and STIS archive, looking for intentional or serendipitous
overlaps with WFPC2 red, or STIS clear images.  Only overlaps
with I$_{814}$ and V$_{606}$ were found. We selected the fields with
exposure time greater than 1000 s. Fields with galactic latitude
$|b|<10\deg$, or E(B-V)$>$0.15 (according to \cite{EXMAPS}) were
discarded to reduce intervening reddening or foreground star
contamination. Fields with HII filaments or with the primary target
extended enough to enter the parallel field of view were discarded as
well.  2) Deep optical R band imaging of two NICMOS parallel fields
was obtained at the ESO-NTT using SUSI2 under superb seeing
($0\farcs5$-$0\farcs6$ measured on the combined image). The 3$\sigma$
limiting magnitude over a 1$\sq''$ area is R=26.8.  3) The optical
counterparts for the HDFS NICMOS flanking fields made available by
Teplitz et al. (1999). We also used the R and I band images of Treu et
al. 1998 (hereafter T98).  4) The NICMOS catalog of part of the HDF
(Thompson et al. 1999) was cross-correlated with the HDF WFPC2 catalog
(Williams et al. 1996). The HDFS NICMOS field (Fruchter et al. 1999,
in preparation) was also considered in the survey, together with the
VLT Science Verification optical data (Renzini 1998).

The 23 fields are listed in Tab~\ref{tab:1} together with the exposure
times in different bands, and the effective area covered. The latter
was computed as the intersection of the IR and optical fields of
view. We ran extensive simulations ($\ge$10,000 per field) in order to
measure the probability of detection of a compact elliptical-like
source as a function of magnitude, and to correct the raw counts for
completeness. At our magnitude limits the completeness was always
higher than 90 \%\footnote{Only in field 1509-1126 the probability
goes down to 77 \% at H=23.5 due to the presence of a bright star.}.

\section{Building the catalog}

\label{sec:data}

The pedestal (e.g., T98) was subtracted from the calibrated data by
using a dedicated software. The images were otherwise reduced in the
standard way. The preliminary H$_{16}$ catalog was obtained by running
SExtractor (\cite{SEx}) on the coadded images. The detected sources
were inspected on the single frames to avoid defects, and
the magnitudes were checked against aperture photometry.  We
considered as resolved all the sources with FWHM as measured with {\sc
imexamine} greater than the FHWM of the stars on the coadded frames
(resolution quality 1). When the signal to noise was high enough, the
dimensions were checked on the single frames.  In fields with no
stars, we considered as resolved all objects with a FWHM greater than
1.5 pixels (resolution quality 2). When WFPC2 images were available
and the objects were detected in optical, we required them to be
resolved also in the optical image. Resolution quality 3 was assigned
to objects with FWHM similar to the stars or of the order of 1.5
pixels if no stars were present. Resolution quality 4 was assigned to
objects more compact than stars or smaller than 1.5 pixels, if no star
were present. Only objects with resolution quality 1 and 2 were
considered in the final sample.

As the main aim of this survey is to provide a morphologically
selected sample of HizECs, we limited the H$_{16}$ band catalog to the
relatively bright sources. The limiting magnitudes listed in Tab.~1
were obtained by cutting each luminosity sorted catalog at the level
where all the sources were detected with a {\sc magerr\_best} SExtractor
parameter smaller than 0.15 (and then rounded to the next brightest
half magnitude).  Unfortunately, the HDFS NICMOS flanking fields have
a relatively shallow optical counterpart and therefore it was
pointless for our purposes to go fainter then H$_{16}$=23.5. The limit
H$_{16}$=24.5 was used for the HDF and the HDFS, where deeper optical
images are available.

After cross-identification of the brightest sources, the H$_{16}$
images were aligned with the other passbands and aperture photometry
was performed for all the resolved objects. We used large apertures in
order to measure total magnitudes. As a double check, SExtractor was
run on the optical images. The two photometries were consistent within
the errors for all the objects in the final catalog.

The objects not detected in the optical were assigned an upper limit
to the luminosity corresponding to the 3$\sigma$ background
fluctuation on a fixed aperture of 0.25 $\sq''$ for the
HST images, and two times the FWHM for the CTIO and NTT images. As the
CTIO images were taken under different seeing conditions, with
different pointings and cameras, we did not combine them but instead we
computed the limiting magnitude from the combined probability of
non-detection. The combined limiting magnitudes are I=25.7 and
R=26.45.  When only the images by Teplitz et al. (1999) were available
the 3 $\sigma$ limiting magnitudes were measured to be I=25.6 and
R=26.05.

The colors were finally corrected for galactic extinction using the
coefficients and the E(B-V) values given by Schlegel, Finkbeiner \&
Davis (1998).

All the resolved objects in the catalog that satisfied the color
selection criteria were morphologically classified independently by
the authors, dividing them into three morphological bins: E/SO, S,
Ir. Only the objects for which we agreed on the classification were
assigned a morphological type.

\section{Results and conclusions}

\label{sec:res}

The observed integrated surface density of HizECs as a function of
magnitude is shown in Fig.~\ref{fig:dens} for the low (hexagons) and
high (triangles) redshift selection criteria. It is worth stressing
that this measurement was performed in random fields whereas many of the
known EROs or high redshift ellipticals are radio selected, or found
in the vicinity of QSOs or AGN, so that their inferred surface
density may be biased.

\placefigure{fig:dens}

Fig.~\ref{fig:dens} also shows the predictions of a class of
models. These are similar to the ones computed by Zepf (1997), and
assume a constant comoving density, passive evolution after a single
burst of star formation at $z=z_f$, and a Schechter local luminosity
function (LF). The parameters for the local LF were taken from Gardner
et al.\ (1997) scaling the total normalization factor to the fraction
of E/S0 (32\%; \cite{GD98}). The predictions of the models are shown
as solid (high redshift) and dotted (low redshift) lines for different
cosmologies and different redshift of formation. In each case we
scaled the normalization factor in order to match the observed counts.
We confirm previous results (Cowie et al.\ 1994; Zepf 1997; Barger et
al.\ 1999; Kodama, Bower \& Bell, 1999; see however Benitez et al.\
1999) that fewer objects are observed than predicted by a passively
evolving model of galaxies formed at very high redshift (5 or 10
depending on cosmology). In addition it can be noticed that: 1) Models
with a very low redshift of formation for the stellar populations of
E/S0s ($z \la 2$, cases (a) and (b) in Fig.~1) do not account
for the observed population of HizECs, as no object red enough is
predicted. 2) Models where a significant fraction of the current E/S0
population was already assembled at $z$=1--2 and the stellar
population was formed at higher redshift (5 or 3) describe the
observed counts well. The exact fraction of objects depends to some
extent on the normalization of the local LF, but the results are
qualitatively unchanged if different E/S0 LFs are used. For example,
using the parameters from Marzke et al.\ (1998), about 25\% of current
E/S0s are needed to fit our data for case (e).  3) If we halve the
local LF, assuming that only Es form in short bursts
of star formation at high redshift, the fraction of progenitors
already assembled at high redshift doubles: the data are consistent
with a scenario where a large fraction of Es form at high redshift and
evolve passively, while the S0s form later or undergo late episodes of
star formation or interactions. Surface photometry can help addressing
this issue. 4) It is possible to break the degeneracy between redshift
of formation and fraction of population by means of two selection
criteria sensitive to slightly different redshift intervals. For
example model (d) can be made to match the broad criterion number
counts, but does not predict any object red enough to satisfy the high
redshift criterion. The same qualitative behavior can be noticed in
model (c) even though the errors are still too large to rule out this
particular model.

It should be noticed that our measurement of HizECs number density may
be considered as a lower limit on the actual number of objects
because: {\it i)} very compact, not resolved, E/S0s drop out of the
sample as they are not resolved; {\it ii)} E/S0s may be rejected by
their morphology if, e.g., they are disturbed by a dwarf companion or
by an interaction.  In addition, single burst models produce the
reddest possible colors for a given $z_{f}$, cosmology, and
metallicity. Therefore also the color selection criteria are tuned to
produce a lower limit.  Unfortunately, the residual possibility of
contamination from regular-looking dusty starbursts prevents us from
pushing the lower limit interpretation of our measurement. A
spectroscopic follow up of the brightest candidates is needed to asses
the magnitude of this contamination, if at all present.

Taking our result at face value, a large fraction of the zero redshift
E/S0 population is not detected as HizECs at z$\ga1$. Possible
interpretations are that they may have been assembled at lower
redshift, may be interacting, or they may be bluer than required by
our color selection criteria due to a more complicated star formation
history with delayed star formation (e.g., Jimenez et al.\ 1999). We
plan to test the prediction of alternate models in paper II.

\acknowledgments

We thank R.~P. van der Marel who developed the Pedestal Extraction
Software; O.~Hainaut who took the NTT images; P.~T.~de Zeeuw for a
careful reading of the manuscript; M.~Carollo for useful discussion.
We used NED, ADS, and the LANL e-print archive.  This
research was partially funded by STScI DDRF grant 82216.

\clearpage
\figcaption[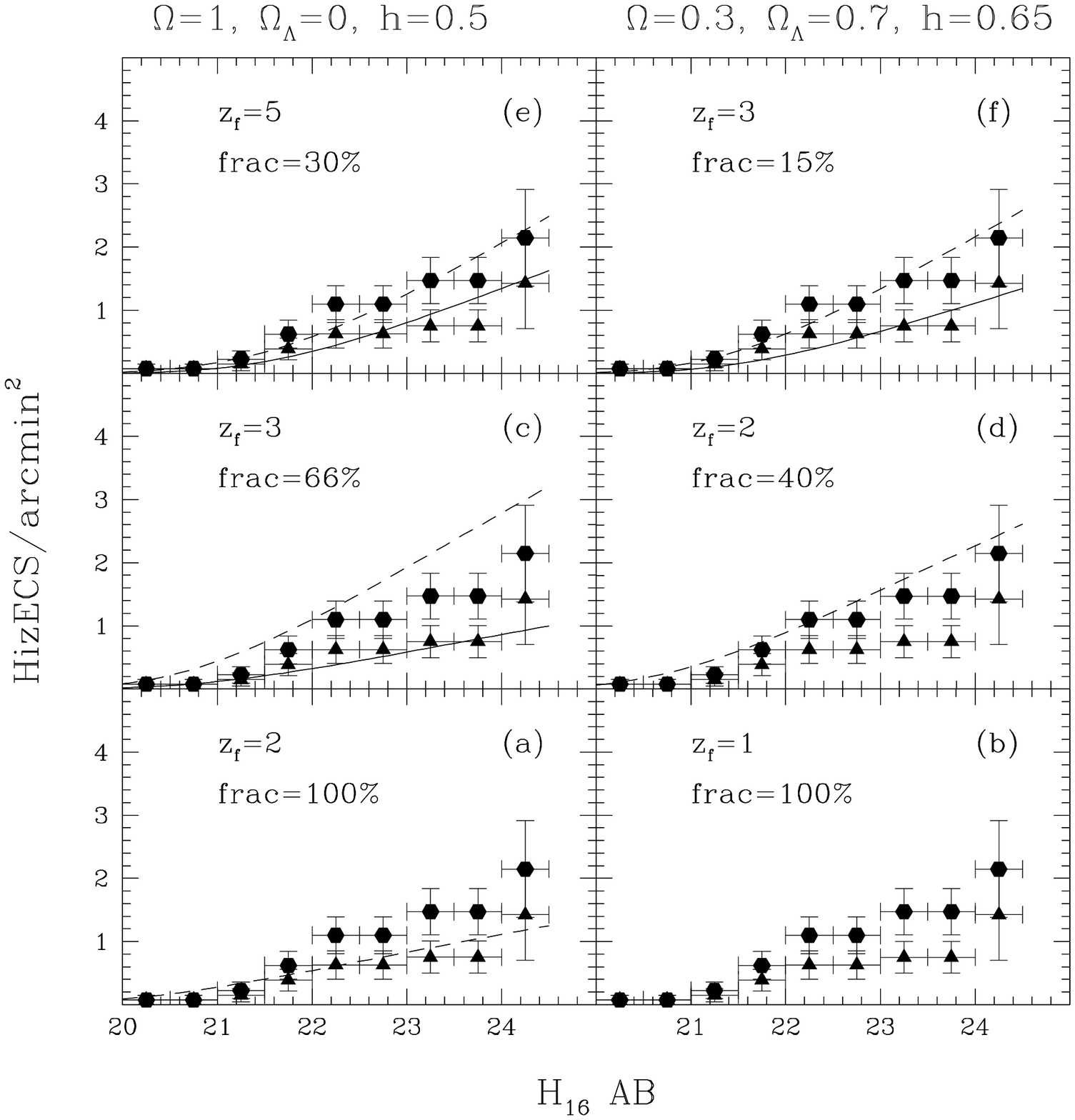]{Integrated number density of HizECS for the low
(hexagons) and high (triangles) redshift selection criteria. Simple
model predictions, assuming the LF from Gardner et al.\ (1998),
constant comoving density and passive evolution are overplotted as
solid (high redshift) and dashed (low redshift) lines (see text for
details). \label{fig:dens}}

\clearpage
 
\begin{deluxetable}{cccccccc}
\footnotesize \tablecaption{Observed fields and exposure times \label{tab:1}}
\tablewidth{0pt}
\tablehead{\colhead{$\alpha$}   & \colhead{$\delta$}   & \colhead{t$_{160}$} & \colhead{t$_{814}$/I} & \colhead{t$_{606}$/R} & 
\colhead{m$_{160}$}   & \colhead{Area} 
}	
\startdata
	  
	   00 17    & +15 43	 & 512    &  -   		& 2200$^a$  & 22   & 0.6778 \nl
	   07 38    & +05 07	 & 1536   &  -	 		& 1800$^a$	 & 23   & 0.7019 \nl
	   09 20    & +69 09	 &  512   &  -	 		& 1500$^a$	 & 22   & 0.6300 \nl
	   10 56    & -03 37	 & 1024	  & 6500$^a$ & 6500$^a$	 & 23   & 0.7212 \nl
	   13 11    & -01 20	 & 1024	  & 2300$^a$ & 1800$^a$	 & 22.5 & 0.1713 \nl
	   13 38    & +70 12	 & 256 	  & 1700$^a$ & 1160$^a$	 & 21.5 & 0.4946 \nl
	   13 38    & +70 13	 & 256 	  & 1700$^a$ & 1160$^a$	 & 21.5 & 0.1330 \nl
	   14 55    & -33 33	 & 2048	  & -	 		& 3600$^b$ & 22.5 & 0.7281 \nl
	   15 09    & -11 26	 & 3071	  & -	 		& 3600$^b$ & 23.5 & 0.6970 \nl
	   21 50    & +28 45	 & 512	  & 8100$^a$ & 7100$^a$	 & 22.5 & 0.6806 \nl
	   22 08    & -19 47	 & 1024	  & -	 		& 1000$^a$	 & 22.5 & 0.2546 \nl
	   22 32    & -60 38	 & 128441 & 10158$^c$  & 7200$^c$   & 24.5   & 0.9569 	\nl
	   22 33    & -60 40	 & 5376	  & 7900$^d$ & 7800$^d$  & 23.5   & 0.6139$\times$7 \nl
	   22 33    & -60 40	 & 5376	  & 5100$^d$  & 4800$^d$   & 23.5   & 0.6110$\times$2 \nl
	   22 32    & -60 43	 & 5824	  & 5100$^d$  & 4800$^d$   & 23.5   & 0.7084$^e$ \nl
	   12 36    & +62 12	 & 131712 & 123600 & 109050 & 24.5   & 0.6601 \nl
\enddata
\tablecomments{The coordinates (hh and mm) are J2000, exposure times
are in seconds, areas are in $\sq$'. The column labeled m$_{160}$
lists the limiting magnitude adopted for the H$_{16}$ catalog. The
flanking fields with the same exposure times have been grouped
together. a) I$_{814}$ and V$_{606}$ HST WFPC2 data; b) ESO NTT SUSI2
c) ESO VLT-UT1 Science Verification. Also VLT-Test Camera U B and V,
STIS CLEAR, and CTIO I, R, V, U, are available. d) R and I taken at
the CTIO 4m Blanco Telescope; e) Total area covered by two fields
partially overlapping. One of the fields has t$_{160}$=7296.}

\end{deluxetable}

\clearpage

\plotone{dens.eps}

\end{document}